\documentclass[aps,twocolumn]{revtex4}
\usepackage{graphicx}
\usepackage{epsfig}
\begin{document}
\bibliographystyle{apsrev}
\title{Extending the WMAP Bound on the Size of the Universe}
\author{Joey Shapiro Key and Neil J. Cornish}
\affiliation{Department of Physics, Montana State University, Bozeman, MT 59717}
\author{David N. Spergel}
\affiliation{Department of Astrophysical Sciences, Princeton University, Princeton, NJ 08544}
\author{Glenn D. Starkman}
\affiliation{Center for Education and Research in Cosmology and Astrophysics,
Department of Physics, Case Western Reserve University, Cleveland, OH 44106--7079 \\
and Astrophysics Department, University of Oxford,  Oxford OX1 3RH, UK}

\begin{abstract}
Clues to the shape of our Universe can be found by searching the CMB for matching circles of
temperature patterns.  A full sky search of the CMB, mapped extremely accurately by NASA's WMAP
satellite, returned no detection of such matching circles and placed a lower bound on the size
of the Universe at 24 Gpc.  This lower bound can be extended by optimally filtering the WMAP power
spectrum. More stringent bounds can be placed on specific candidate topologies by using a
a combination statistic.  We use optimal filtering and the combination
statistic to rule out the infamous ``soccer ball universe'' model.
\end{abstract}

\maketitle

\section{Introduction}

What is the shape of space?  While this question may have once seemed more philosophical 
than scientific, modern cosmology has the chance to answer it using the oldest observable 
light in the Universe, the Cosmic Microwave Background radiation (CMB).  NASA's Wilkinson 
Microwave Anisotropy Probe (WMAP) has made a detailed map of the CMB sky which has
been used to provide answers to many age-old questions about the nature of the Universe~\cite{wmap}. 

While it is certainly possible that the Universe extends infinitely in each spatial 
direction, many physicists and philosophers are uncomfortable with the notion of a
universe that is infinite in extent. It is possible instead that our three dimensional Universe has 
a finite volume without having an edge, just as the two dimensional surface of the Earth 
is finite but has no edge.  In such a universe, it is possible that a straight path in one 
direction could eventually lead back to where it started.  For a short enough closed path, 
we expect to be able to detect an observational signature revealing the specific topology 
of our Universe~\cite{stark}.

\section{Geometry and Topology}\label{topology}

An important question answered by the WMAP mission is that of the curvature of space.  
The matter and energy density of the Universe indicate that space is very nearly flat.
The WMAP data point to a universe with a total energy density within 2\% of
critical~\cite{spergel}.  This means that even if 
space in not quite flat, the radius of curvature of the Universe is at least of order the 
size of the observable Universe, and space can be considered to be nearly flat.
 
The WMAP sky also provides clues about the topology of the Universe.  Observable 
topologies would have a characteristic fingerprint in the CMB sky.  The set of possible 
topologies for our Universe is determined by the curvature of space.  In a flat Universe, 
the allowable topologies are restricted to a set of eighteen possibilities.  It has been 
shown that a nearly flat Universe would have an observational fingerprint very similar to 
that of an exactly flat Universe~\cite{jeff2,reza}.  Cornish, Spergel, and Starkman~\cite{stark} 
have described the CMB signature revealing the shape of space as ``circles in the sky''.

\section{Circles in the Sky}\label{circles}

\begin{figure}[h]
\includegraphics[angle=0,width=0.4\textwidth]{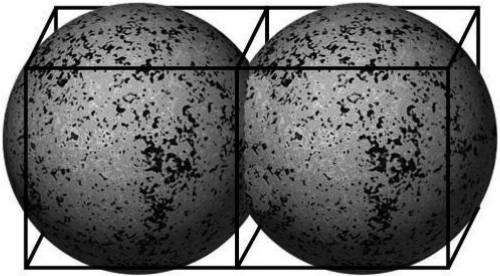}
\caption{\label{spheres_fig}In a multiply connected universe, all of space must be tiled with 
copies of the fundamental domain.  Here the fundamental domain is a cube, and associated with
each domain is a copy of the last scattering surface.
The intersections are physically the same place in space, producing 
matching circles of temperature patterns in two different directions as observed from the 
Earth.}
\end{figure}

Circles in the CMB revealing the topology of the Universe could only be observed for a universe 
that admits sufficiently short closed paths.  If our Universe is multiply connected and the fundamental 
domain fits inside the observable Universe, we expect to see the characteristic signature of 
the topology in the CMB.  In a Universe with a non-trivial topology, there exist multiple 
copies of objects, one in each copy of the fundamental domain.  A copy of our own galaxy, 
solar system, and planet could possibly be observed many light-years away.  Surrounding 
each copy of Earth would be a copy of the last scattering surface - a 2-sphere at redshift
$z\simeq 1100$. Since in a homogeneous and isotropic geometry the intersection of two 2-spheres 
is a circle, and these circles in the CMB are physically the same place in space, to the
extent that the CMB measures the local gravitational potential, density and temperature on
the last scattering surface, we expect 
to find matching circles of temperature patterns in the CMB when looking in two different 
directions in the sky (Figure~\ref{spheres_fig}). 
 
The search for these matching circles of temperature patterns in the WMAP data is a huge task.  
The resolution of the WMAP satellite provides over three million independent data points on 
the sky.  Each of these points can be considered a circle center, with the circle radius 
ranging between $0$ and $90^{\circ}$.  Once the temperature pattern around a given circle is found, it must 
be compared to every other circle in the sky with the same angular radius, and the matching 
statistic calculated for the pair.  The phasing of each circle pair must also be considered, 
ranging from $0$ to $360^{\circ}$, since the start of the comparison can begin anywhere along the second 
circle.  This full search is time consuming, and is still in progress.

A more efficient search can be employed when considering the WMAP result of nearly zero spatial 
curvature for the Universe.  In a flat universe, all eighteen possible topologies produce some circle 
pairs that are back-to-back (though the smallest circles may not be back-to-back).  That is, their
circle centers are $180^{\circ}$ apart.  It can also be shown that for a large radius of curvature
compared to the radius of the last scattering surface, non-trivial topologies 
with a positive curvature produce nearly back-to-back circle pairs~\cite{jeff2,reza}.
A shorter search can thus 
be used on the WMAP data, where each circle is compared only to its back-to-back pair.  Similarly, 
an almost back-to-back pair search can be implemented, looking only for matching circles between 
$170^{\circ}$ and $180^{\circ}$ from the first circle center. This search has been completed, and
no statistically significant matches were found~\cite{cornish}.

\section{The $S$ Statistic}\label{S}

A matching statistic is needed to define the correlation between the temperature patterns 
around two circles with the same angular radius.  In Ref.~\cite{cornish}, the $S$ statistic for circle 
comparison is defined, 
 
\begin{equation}\label{stat}
S_{ij}(\alpha,\beta) = \frac{2\sum_{m} m T_{im}(\alpha)T_{jm}^{*}(\alpha) e^{-im\beta}}{\sum_{n} n
\left[ \left\vert  T_{in}(\alpha) \right\vert ^2 
+ \left\vert T_{jn}(\alpha)\right\vert ^2 \right]}.
\end{equation}

Here $T_{im}$ denotes the $m^{\rm th}$ harmonic of the temperature pattern around the $i^{\rm th}$
circle. The parameters $\alpha$ and $\beta$ denote the angular radius and the phase offset between
the circles, respectively. The $S$ statistic is normalized with a perfect match yielding an $S$ 
value of 1.  The comparison is done in Fourier space, in order to include appropriate weighting of different 
angular scales.  The $m$ weighting factor on circles plays the same role as the usual $\ell (2\ell +1)$
weighting that appears in the full angular power spectrum: these factors take into account the
numbers of degrees of freedom per mode~\cite{gorski}. If the $m$ weighting is neglected
large angular scales dominate and completely drown out small-scale fluctuations.  
A limited search for matching circles in the CMB sky has also been performed in Ref.~\cite{roukema}.  There, 
the matching statistic of~\cite{stark} was used; one that does not include the appropriate $m$ weighting 
factor:  
 
\begin{eqnarray}\label{no_m}
S &\equiv& \frac{\left\langle 2\left( \frac{\delta T}{T}(\alpha,\phi)\right)_{i}
\left( \frac{\delta T}{T}(\alpha,\phi+\beta)\right)_{j}\right\rangle}
{\left\langle \left( \frac{\delta T}{T}\right)_{i}^{2} + \left( \frac{\delta T}{T}\right)_{j}^{2}\right\rangle}.
\nonumber \\
&=& \frac{2\sum_{m} T_{im}(\alpha)T_{jm}^{*}(\alpha) e^{-im\beta}}{\sum_{n} 
\left[ \left\vert T_{in}(\alpha) \right\vert ^2 + \left\vert T_{jn}(\alpha)\right\vert ^2 \right]}.
\end{eqnarray}
 
\begin{figure}[h]
\includegraphics[angle=0,width=0.45\textwidth]{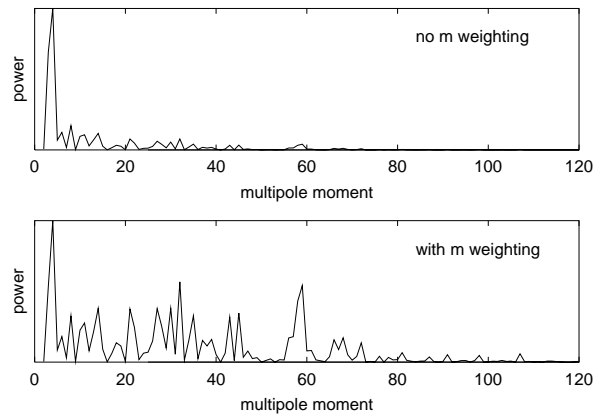}
\caption{\label{weighting_fig}The Fourier coefficients of $\delta T$, as a function of multipole moment $m$,
around a circle with no $m$ weighting 
(top) and including $m$ weighting (bottom). }
\end{figure}

Here the brackets denote integration along the circle.  This alternative definition of the $S$ 
statistic can be calculated in position space, but the consequence of disregarding the weighting 
factor is that large scale temperature variations swamp any small scale features. The importance
of the $m$ weighting is apparent in Figure~\ref{weighting_fig}. 
 
\section{False Positive Calculation}\label{false}

An essential aspect of our search is the calculation of the 
threshold for a significant match.  We expect this false positive line to be higher for circles 
with smaller angular radius, since these circles have fewer pixels used for comparison and thus 
have a greater chance for random matching.  It has been shown in Ref.~\cite{cornish} that the level for a 
positive detection can be calculated using 
 
\begin{equation}\label{fp}
S_{\rm max}^{\rm fp}(\alpha) \simeq \langle S^{2}\rangle^{\frac{1}{2}}\sqrt{2 \ln \left(\frac{N_{\rm search}(\alpha)}
{2\sqrt{\pi \ln ( N_{\rm search}(\alpha))}} \right) }.
\end{equation}
 
Here $N_{\rm search}$ is the size of the search space, counting the number of circle comparisons made in the 
search at each angular radius, $\alpha$.  
The false positive threshold is sensitive to $N_{\rm search}$, increasing for 
larger searches.  The expected value of $S$ for random skies can be estimated from first principles,
but it is better to numerically measure the expectation value from the data.
The expectation value of $S$, ${\left\langle S^{2} \right\rangle}^{\frac{1}{2}}$, 
and the size of the search space, $N_{\rm search}$, 
follow from the implementation of the specific search being performed.  When looking at the results 
from our searches, we plot a false positive threshold for which we expect fewer than 1 in 100 
random skies to produce an $S$ value above the line. 

Calculating the correct value of $N_{\rm search}$ involves determining the number of completely independent 
possible searches.  For example, a sky with $2^{\circ}$ smoothing has less independent pixels than a sky 
with $1^{\circ}$ smoothing and thus has fewer possible independent circle pairs to be searched.  The full 
$N_{\rm search}$ value for a search should take into account 
the number of independent circle radii ($\alpha$) searched at a given sky resolution.  This was not included 
in Ref.~\cite{cornish}, but we include it here by multiplying the $N_{\rm search}$ value at each $\alpha$
calculated by the search code by a factor equal to the number of independent radii searched.
The false positive level must be considered for any claim of detection, and is used to 
place the lower bound on the size of the Universe for no circle pair detections.  

\section{The WMAP Bound}\label{bound}

\begin{figure}[t]
\includegraphics[angle=0,width=0.5\textwidth]{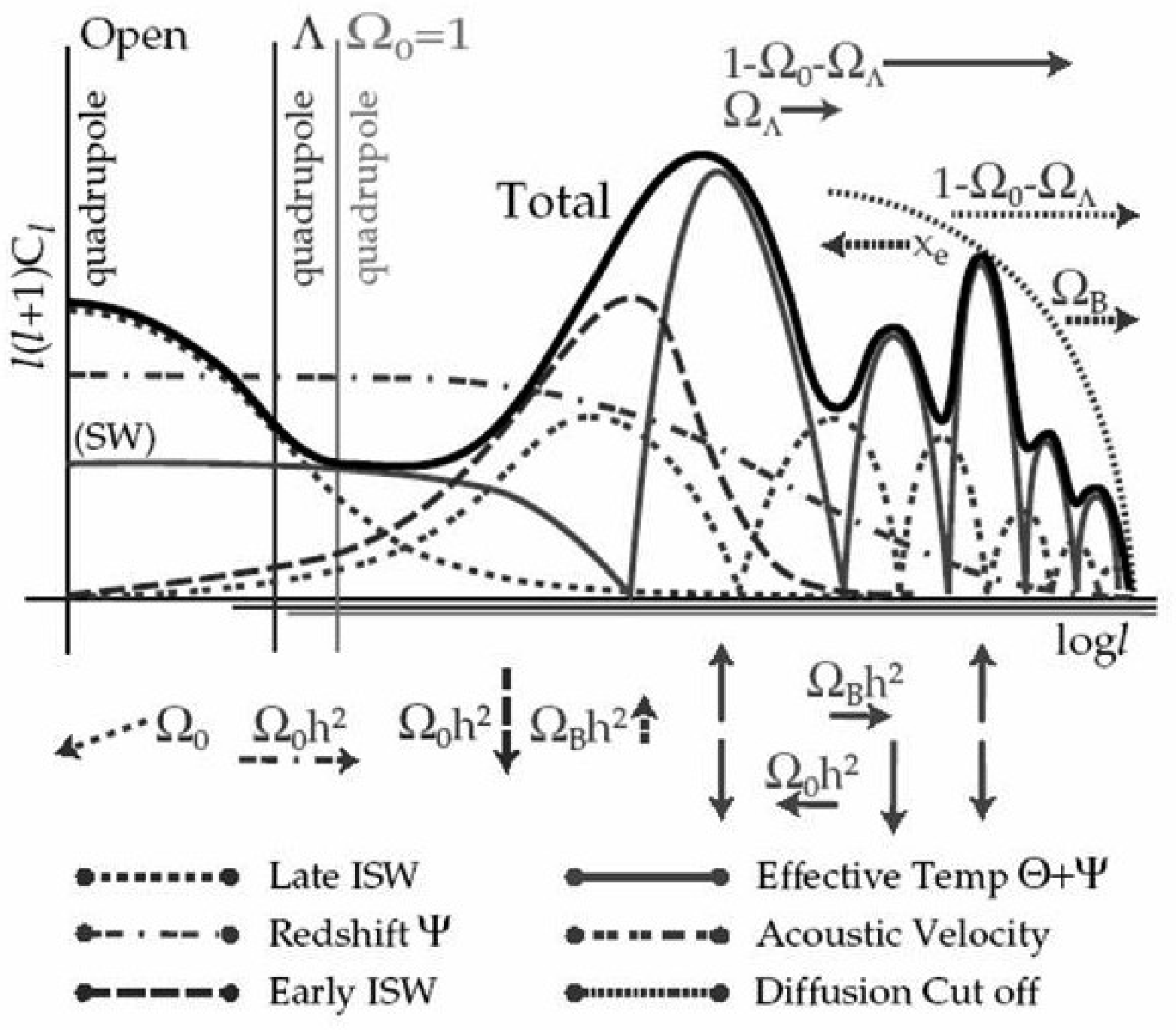}
\caption{\label{cmb_fig} Contributions to the power spectrum of the CMB.  The effective 
temperature around 
circles can be correlated due to topology, but is degraded by ISW effects.  The acoustic 
velocity also degrades correlation for small circles.  
(Reproduced with permission from \emph{Nature}, Hu, Sugiyama, \& Silk (1997)) }
\end{figure} 

The full sky search for back-to-back and almost back-to-back matching circles was performed in
Ref.~\cite{cornish}, and it is apparent in Figures 3 and 4 of that paper that
no statistically significant matches were found.
 
A simulation of the CMB has been produced~\cite{cornish} that includes all relevant physics,
using model parameters that give a good match to the real CMB power spectrum, but with a non-trivial
topology built in.  This 3-torus universe 
with a fundamental domain that fits inside the sphere of the CMB is used to test the search codes.  
The results of the back-to-back search on the simulated 3-torus universe indeed show peaks in the 
matching statistic, indicating matched circle pairs (Ref.~\cite{cornish}, Figure 1).  The fact that 
the results of the same back-to-back and nearly back-to-back searches of the WMAP data show no 
such peaks can be used to place a lower limit on the size of our Universe.  The intersection of 
the false positive threshold and the matching statistic peaks found for a simulated 3-torus 
universe defines this lower limit, since it indicates that the smallest circle pairs we could 
expect to detect would be approximately $20^{\circ}$. 

\begin{figure}[t]
\includegraphics[angle=0,width=0.3\textwidth]{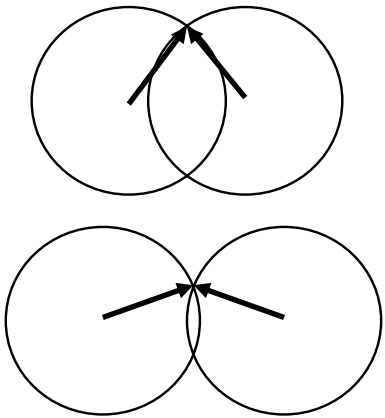}
\caption{\label{intersect_fig}Here a circle represents a cross section of the 2-sphere of the CMB.  
The point of 
intersection thus represents a point along a circle in the sky, viewed from two different copies 
of the Earth.  It can be seen that small circles have an almost anti-parallel velocity contribution,
$\hat{n} \cdot \vec{v}$
(top), while large circles have an almost parallel velocity term (bottom). }
\end{figure}

For decreasing $\alpha$, a downward trend can be seen in the 3-torus simulation results for the height 
of the peak indicating a match. Photons from the last scattering surface travel toward Earth, carrying
with them an imprint of the conditions at their point of origin.  It is this effective temperature of the 
CMB photons that should match along a circle.  However, two matching photons traveling different 
paths toward Earth will encounter different line of sight effects, degrading the temperature 
match.  The combined line of sight effects on CMB photons can be understood by considering the 
physical contributions to the total power spectrum of the CMB (see Figure~\ref{cmb_fig}).  The Integrated 
Sachs-Wolfe effect (ISW) degrades any possible match, while the contribution by the acoustic 
velocity term depends on the size of the circle pair.  The velocity term is given by
$\hat{n} \cdot \vec{v}$, where $\hat{n}$ is a unit vector along the line of sight and $\vec{v}$
is the velocity vector of the plasma at the surface of last scatter. 
For large circles, this term is correlated, while for small circles this term becomes
increasingly anti-correlated as the circles get smaller.
(see Figure~\ref{intersect_fig}).  This velocity 
term thus degrades the matching statistic value for small circles and explains the downward 
trend of peaks for small $\alpha$.

The null result of the search for matching circles in the CMB indicates that the fundamental 
domain of the Universe must be at least on order the size of the surface of last scatter.  The fact 
that the false positive line intersects the maximum peaks expected for the matching statistic at 
$\alpha=20^{\circ}$ means that the fundamental domain must be big enough such that only circles 
smaller than 
$20^{\circ}$ could be produced by intersections of copies of the CMB.  This, along with the best fit values
for the other cosmological parameters, places a lower bound on the 
size of the Universe at 24 Gpc.

\section{Extending the Bound}\label{extend}

There are a couple of techniques that can be applied in an attempt to extend this bound beyond 
24 Gpc (or possibly detect the topology of the Universe, lurking just below the current false 
positive threshold).  Since we are interested in raising $S$ values for small circles where 
it dips below the detection line, we can filter out from the CMB power spectrum the terms that 
degrade matches for small circles.  This requires filtering out both the ISW effect on large
angular scales and the acoustic velocity term on smaller angular scales.  We thus multiply the
WMAP power spectrum by a filter that includes only 
the regions of the power spectrum where the effective temperature dominates all other terms.   

\begin{figure}[t]
\includegraphics[angle=0,width=0.45\textwidth]{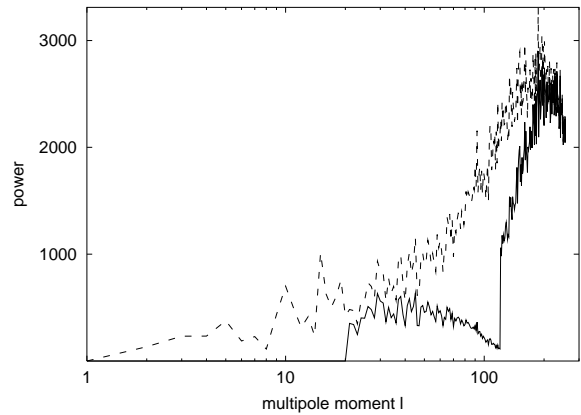}
\caption{\label{power_fig} The unfiltered simulated power spectrum (dotted line) and the power
spectrum with 
regions dominated by either ISW or acoustic velocity effects filtered out (solid line).}
\end{figure}

The basic shape of our filter comes from the physical theory described in Figure~\ref{cmb_fig}.  
We then search 
for the best filter on the 3-torus simulation, varying the location, width, and steepness of the 
filter windows.  Our best simulated filtered power spectrum is seen in Figure~\ref{power_fig}.   
 
\begin{figure}[h]
\includegraphics[angle=0,width=0.4\textwidth]{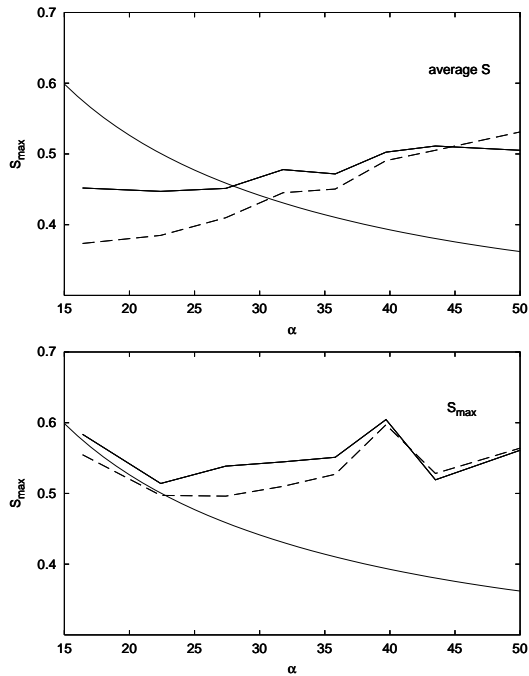}
\caption{\label{filter_fig}The dashed lines give the $S$ values for the unfiltered simulation,
while the solid lines give the $S$ values when the power spectrum filter is included.
Top: The average values of the $S$ statistic as a function of 
angular radius.  Bottom:  
The maximum values of $S$ as a function of angular radius.  $S_{max}$ increases by $\sim 10\%$ for 
small circles 
with filtering, but decreases for large circles, as expected.  Plotted with the calculated false 
positive line.}
\end{figure}

The filtering indeed raises the $S$ values for small circles in our simulated 3-torus universe and 
lowers the $S$ values for large circles, as expected when filtering out the velocity term that 
contributes to raising $S$ for large angular radius pairs (Figure~\ref{filter_fig}).  

In general, the false positive line depends on the filter being used.  A filter that keeps 
only the low $l$ power, for instance, would result in a higher false positive line since there are 
essentially less significant pixels without the high $l$ values, and thus a greater chance of a 
false match.  Our power spectrum filter keeps most of the high $l$ values, especially near the 
first acoustic peak, and thus does not significantly affect the false positive line.

\begin{figure}[t]
\includegraphics[angle=0,width=0.4\textwidth]{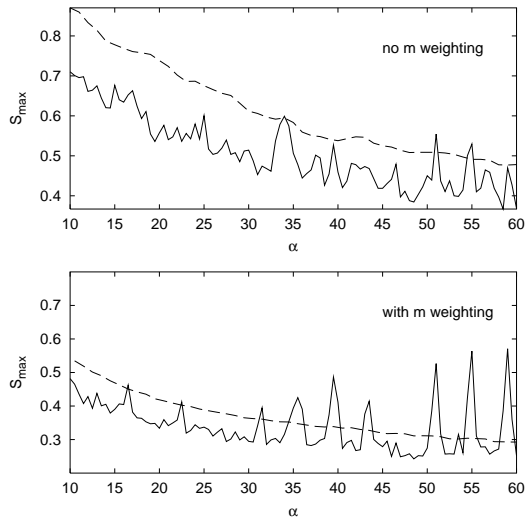}
\caption{\label{peaks_fig} A back-to-back circle search is performed on our 3-torus
simulation with $1^\circ$ degree smoothing.  The dashed line shows the results for a
search with the matching statistic
calculated without the necessary $m$ weighting, while the solid line indicates a search
including $m$ weighting. The threshold line for detection for each search is also plotted.}
\end{figure}

Another choice of filter is the use of $m$ weighting, as discussed in Section \ref{S}.  The $S$ 
statistic calculation neglecting $m$ weighting results in higher $S$ values for circle pair
comparisons at all angular radii, raises the false positive level, and degrades significant peaks.
These effects are illustrated in Figure~\ref{peaks_fig}, showing the results of a back-to-back circle
search of a 3-torus CMB simulation when $m$ weighting is neglected (dashed line) and included 
(solid line).  It can be seen that the search without $m$ weighting emphasizes large angular
scales, and thus only finds matching circles with $\alpha > 45^{\circ}$.

Several recent papers~\cite{aurich,steiner,then} have called into question the efficacy of the
matched circle test. In each
instance the negative conclusions can be traced to the use of the unweighted $S$ statistic. As our
searches of realistic simulated skies have shown, the properly weighted matching statistic provides
a powerful tool for probing the topology of the universe.

\section{Combination Statistic}\label{combo}

Another technique for probing below the limit of $20^{\circ}$ circle pairs is employing a directed 
search for a specific topology.  Such a search can increase the $S$ value for small circles by stringing 
together several sets of circles in an expected configuration to create an effectively 
larger radius circle with the increased number of pixels.  For a cubic 3-torus universe, for 
example, we expect to find at least three sets of circle pairs corresponding to the matching
faces of the fundamental domain and we can predict the relative orientations of each matching set.   

The combination search for this example would consist of choosing a circle center and circle radius
and comparing the pixels around this circle to the circle exactly opposite it in the sky.  The search then 
chooses an axis for a second set of back-to-back circles with the same angular radius. The location
of the final circle pair is fixed once the axis has been chosen. The pixels along all 
three pairs are strung together to calculate the total matching statistic for the whole set.  
Since there are more pixels for the whole set than for one pair of circles, there 
is a smaller chance for a random match.

This combination statistic can be tested using our simulated 3-torus universe.  There are 
matching circles with angular radius of about $16^{\circ}$ in our simulation.  These pairs cannot be 
detected in the single pair back-to-back circle search because the false positive threshold 
is above the peak of their matching statistic. 

Our 3-torus simulation has a fundamental domain such that several copies fit within the surface of
last scatter. There are thus more than three sets of circles at $16^{\circ}$, and our combination 
statistic is calculated by stringing together the pixels along four sets of circles.

\begin{figure}[t]
\includegraphics[angle=270,width=0.45\textwidth]{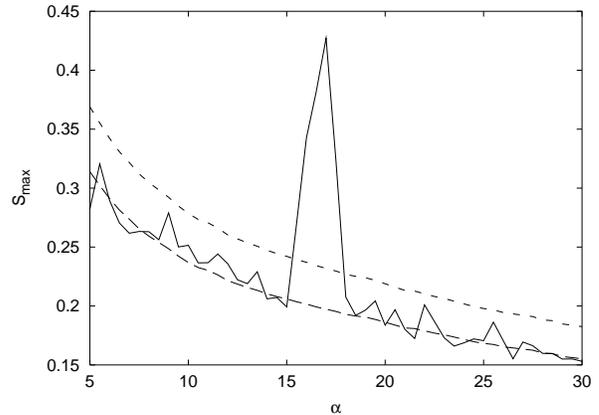}
\caption{\label{combo_fig} The combination statistic calculated for our 3-torus simulation using
the optimal filter (solid line). The upper dashed 
line gives the threshold for less than 1 in 100 random skies producing an $S$ value above the line.}
\end{figure}  
 
It can be seen in Figure~\ref{combo_fig} that the search for the combination statistic in our 
simulated 3-torus universe raises the $S$ value above the false positive line, and we can thus make 
a statistically significant detection.  The 3-torus combination statistic can be applied to the 
real WMAP data to rule out (or detect) sets of matching circles smaller than the $20^{\circ}$ cutoff of 
the single circle pair search. 

It is crucial to calculate the false positive line correctly for the combination search. While 
the combination statistic lowers the false positive level for small circles by creating 
effectively larger circles, it must be remembered that the combination statistic requires a 
larger search than looking for back-to-back circles.  There is an extra degree of freedom 
for the orientation of the fundamental domain once the first set of circles is identified.
The false positive level for the combination search can be estimated from the standard
pair-wise search as follows. The RMS value for $S_{\rm combo}$
is given by $\langle S_{\rm combo}^2 \rangle^{1/2} = \langle S_{\rm pair}^2 \rangle^{1/2}/n^{1/2}$,
where $n$ is the number of circle pairs used in the combination statistic. The number of circle
comparisons in the combination search, $N_{\rm search}^{\rm combo}$ scales as $N_{\rm search}^{(d+1)/d}$,
where $d$ is the number of degrees of freedom in the pair-wise search. Combining these observations
in (\ref{fp}) yields
\begin{equation}\label{fp_combo}
S_{\rm max-combo}^{\rm fp}(\alpha) \simeq \left(\frac{d+1}{nd}\right)^{1/2} S_{\rm max-pair}^{\rm fp}(\alpha) \, .
\end{equation}
We can also estimate the combination statistic level for matching circles in terms of the pairwise values.
The combination statistic for matching circles, $S_{\rm combo}^{\rm match}$,
is the weighted average of the pair-wise statistic for matching circles, $S_{\rm pair}^{\rm match}$,
and since the total power around each circle is roughly equal, we have $S_{\rm combo}^{\rm match} \simeq
\bar{S}_{\rm pair}^{\rm match}$, where the bar denotes the average. These estimates for the false
positive and match level for the combination searches were found to be accurate predictors of the
numerical results. Extrapolating from the values of $\bar{S}_{\rm pair}^{\rm match}$ seen in the upper panel
of Figure~\ref{filter_fig} and the values of $S_{\rm pair}^{\rm match}$ seen in the lower panel of
Figure~\ref{peaks_fig} suggests that a combination search using 4 of more circle pairs should be able to
detect circles with radii as small as $5^{\circ}$. 

\section{Poincar\'{e} Dodecahedral Space}\label{dodec}

\begin{figure}[t]
\includegraphics[angle=0,width=0.45\textwidth]{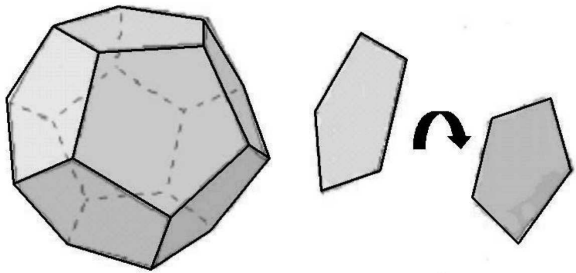}
\caption{\label{dodec_fig}The Poincar\'{e} dodecahedron is a 12-sided polygon with identical faces.  
This shape 
can tile space with no gaps only in a positively curved universe.  The face identifications 
include a twist of $\pm 36^{\circ}$.}
\end{figure}

	In a universe with positive curvature, the faces of a dodecahedron can be identified 
to form what is known as Poincar\'{e} Dodecahedral Space (the ``soccer ball universe'').  
The face identifications for a dodecahedron involve a twist  of $\pm 36^\circ$ to insure the matching
up of the vertices of the face pair (Figure~\ref{dodec_fig}). It has been claimed that the power spectrum
expected for a universe with such a topology closely matches the power spectrum found 
for our Universe by WMAP~\cite{weeks}.
 
\section{Claim of a Non-Trivial Topology}\label{claim}
 
A claim has been made that six sets of matching circles in the CMB have been found, 
indicating that the Universe has the topology of the Poincar\'{e} Dodecahedron~\cite{roukema}.  The claim, 
however, is not accompanied by a statistical analysis of the results, and cannot be assessed
until this is completed.

\begin{figure}[h!]
\includegraphics[angle=270,width=0.45\textwidth]{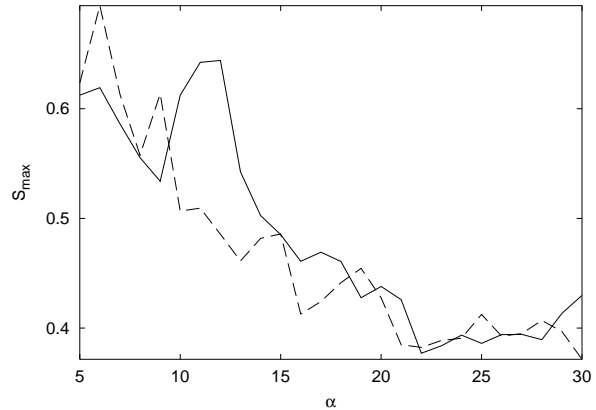}
\caption{\label{alpha_fig} Our combination statistic plotted for circle comparisons with a 
$-36^{\circ}$ twist 
(solid line) and a $+36^{\circ}$ twist (dashed line).  Including $2^{\circ}$ smoothing, cutting out the 
galaxy contamination, and neglecting the $m$ weighting factor reproduces the peak 
found in Ref.~\cite{roukema}.}
\end{figure}

In Ref.~\cite{roukema}, the combination statistic for the expected six sets of matching circles of a Poincar\'{e} 
dodecahedron has been computed for a range of circle radii $\alpha$ with a peak appearing for circles 
around $11^{\circ}$.  
The peak only appears when the $-36^{\circ}$ twist is used when identifying faces of the dodecahedron 
(c.f. Ref.~\cite{roukema}, Figure 4).  This peak is then compared to the same search using the 
$+36^{\circ}$ twist for face identifications (c.f. Ref.~\cite{roukema}, Figure 5), where there
appears to be no peak.  It is expected that the 
peak would appear for only one twist value if this is an indication of the physical topology 
of the Universe.   

\begin{figure}[t]
\includegraphics[angle=0,width=0.45\textwidth]{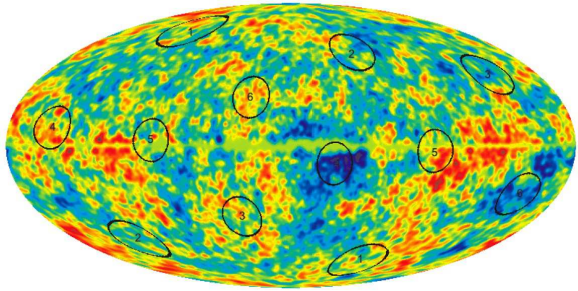}
\caption{\label{location_fig} The locations of the six sets of circle pairs in the CMB sky 
identified in Ref.~\cite{roukema} and found by our combination statistic search.}
\end{figure}

Figure~\ref{alpha_fig} shows that we were able to find the same peak in the matching statistic
around $11^{\circ}$ using a combination statistic. In order to reproduce the peak, we used the 
$S$ statistic given in Equation (\ref{stat}), a galaxy cut, and the $2^{\circ}$ smoothing used in 
Ref.~\cite{roukema}.  The locations of the six pairs in the CMB sky, shown in Figure~\ref{location_fig},
also matches those found in Ref.~\cite{roukema}. 
 
\begin{figure}[t]
\includegraphics[angle=0,width=0.4\textwidth]{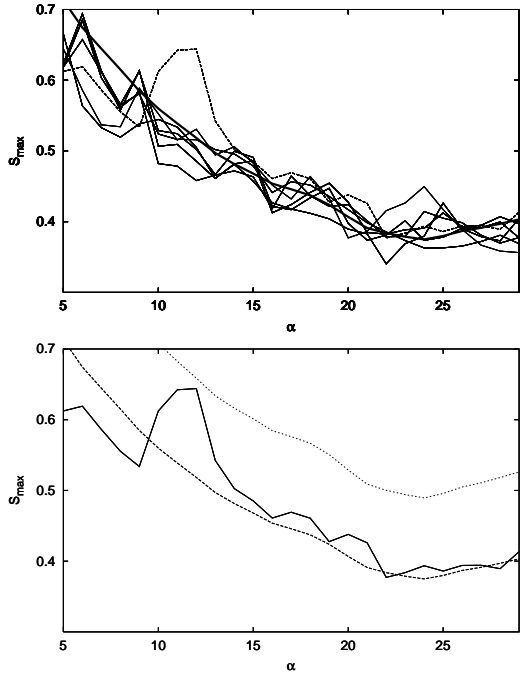}
\caption{\label{scramble_fig}  
Top:  The combination search for the dodecahedral topology performed with several different
twist values.  The expected level for $S$ is plotted in bold.
Bottom:  The possible detection with the $-36^{\circ}$ twist (solid line) is plotted with the appropriate 
expected level for $S$ (lower dashed line) and the threshold for fewer than 1 in 100 random skies 
producing an $S$ value above the line (upper dotted line) }
\end{figure}
 
For proper analysis of the results, the false positive line must be calculated correctly.
The results from the $+36^\circ$ twist search do not give the actual false positive 
line for detection.  To get a better idea of the expected peak levels for a universe with 
a trivial topology, the WMAP data can be randomly scrambled.  This preserves the power 
spectrum and distribution of temperature values while destroying any matching circles that 
would give an indication of topology.  The same search can then be performed on many of these 
scrambled skies and the results plotted together to show the expected random peak levels for 
a trivial topology.  In the case of the position space search used in Ref.~\cite{roukema}, the WMAP
sky cannot be randomly scrambled while preserving both the power spectrum and the galaxy cut.
To get a reliable measure for the expected $S$ values for the combination search we instead use various twists 
other than $\pm 36^\circ$ when identifying circle pairs.  The results from the search on these random twists
is plotted along with the possible detection in Figure~\ref{scramble_fig}.
 
\begin{figure}[t]
\includegraphics[angle=0,width=0.4\textwidth]{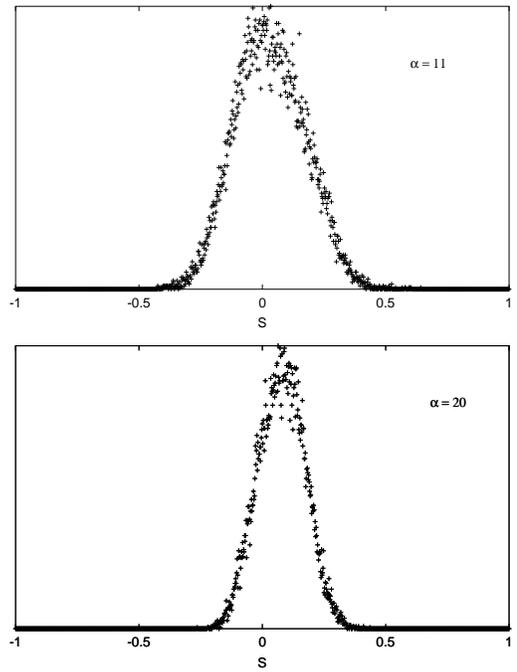}
\caption{\label{gauss_fig} The distribution of $S$ values calculated at $\alpha = 11^{\circ}$ (top)
and $\alpha = 20^{\circ}$ (bottom).  Both have a non-zero mean due to a DC component of the temperature
values around circle pairs.}
\end{figure}

\begin{figure}[t!]
\includegraphics[angle=270,width=0.45\textwidth]{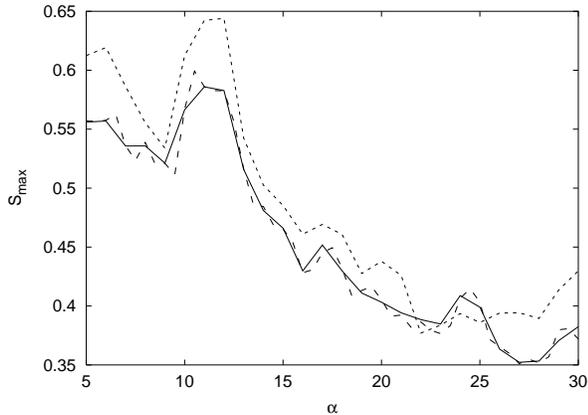}
\caption{\label{fourier_fig} The search in Fourier space with no $m$ weighting 
(solid line) matches the search 
in position space with no galaxy cut and no $m$ weighting (dashed line).  Including the galaxy cut 
raises the peak (dotted line). }
\end{figure}

It can be seen in Ref.~\cite{roukema}, Figures 13 through 18, that two sets of circles 
include points 
that run through the plane of our galaxy.  The CMB data for this portion of the sky is contaminated 
by the radiation from our galaxy. While it is possible to filter out much of this contamination,
a more conservative approach is to perform a galaxy cut.
The galaxy cut is easily done in position space, but is not quite as easily included 
in a Fourier space search.  Including the galaxy cut raises the peak for matching circles, since 
it cuts out uncorrelated points, but it also raises the false positive level, since there are 
less data points to compare.  It can be seen in Figure~\ref{fourier_fig} that the unweighted 
Fourier and 
position space searches produce similar peaks and that the galaxy cut in position space raises 
the peak as expected.  While Parceval's theorem states that the position space search with no galaxy cut 
and the Fourier search with no $m$ weighting should produce exactly the same result, numerical error
leads to a slight difference in the results.

\begin{figure}[t!]
\includegraphics[angle=0,width=0.4\textwidth]{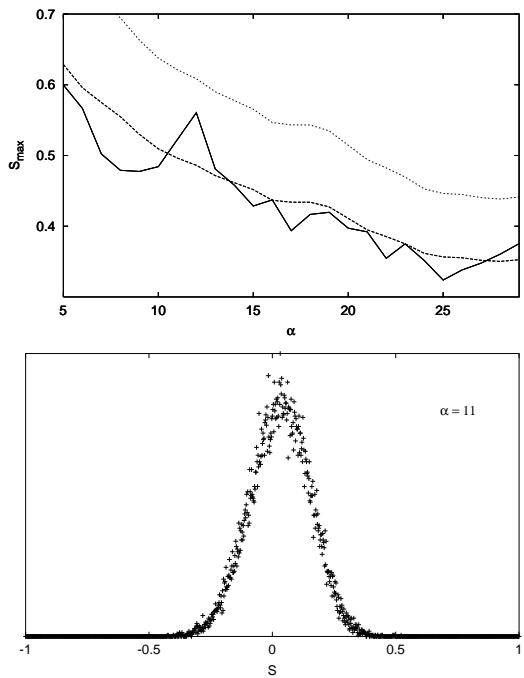}
\caption{\label{dc_fig} Removing the DC contribution (average value) when comparing circle pairs 
shifts the distribution of $S$ values toward zero mean and indicates no peak at $11^{\circ}$.}
\end{figure}

We expect that the $S$ values calculated for all of the sets of circle pairs compared at a give $\alpha$
should form a Gaussian distribution.  Even if there is an outlier value of $S_{max}$ indicating
the topology of the Universe, this one value should not significantly skew the distribution of
all of the non-detections at the same angular radius.  Two distributions of calculated $S$ values for
the $-36^{\circ}$ twist search can be seen in Figure~\ref{gauss_fig}.  The non-zero mean of the 
distributions indicates an average value for the distribution above zero, and thus a DC contribution to 
the temperature values around each circle pair.  This 
contribution can be subtracted to show the true comparison of temperature variation.  Simply removing the 
DC contribution reduces the peak in the $S$ values of the dodecahedron combination search 
(Figure~\ref{dc_fig}, top panel).  The distribution of $S$ values is shifted toward zero mean by explicitly 
removing the average temperature values, but the mean is not exactly zero.  This is due in part to
numerical error.  The $S$ statistic calculated using $m$ weighting (Equation (\ref{stat})) automatically 
takes care of this average offset.

\begin{figure}[t]
\includegraphics[angle=270,width=0.45\textwidth]{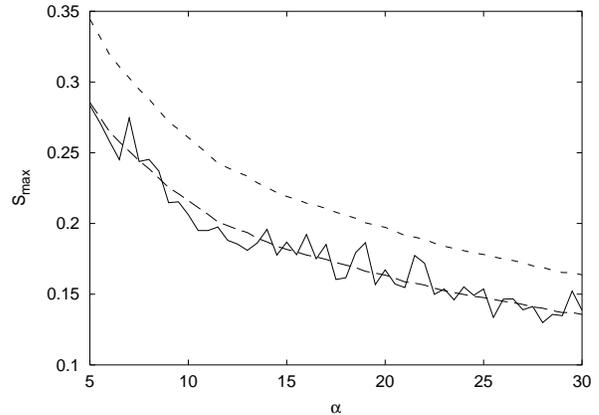}
\caption{\label{final_fig}  Our results for the Poincar\'{e} Dodecahedron combination search with 
optimal filtering. The dashed line indicates the detection threshold.}
\end{figure}

Figure~\ref{final_fig} shows our final analysis, using the Poincar\'{e} combination search with 
the appropriate $m$ weighting and plotted with the threshold line for detection.  The appropriate
analysis yields no 
significant peak at $11^{\circ}$.  It is thus clear that when the the proper matching statistic is
used, and the expected level of false positives is calculated, that the detection claimed in
Ref.\cite{roukema} does not hold up. It seems that the observed peak is most 
likely due to overemphasis of large angular scales by neglecting $m$ weighting, resulting in 
searching for matching circles in a sky with large patches of constant temperature.  Such a 
sky would indeed produce randomly matched small circles.  
  
It should also be noted that a search for matching circles has been performed in Ref.~\cite{aurich} and
Ref.~\cite{steiner}.  The authors found that their circle matching 
statistic was degraded considerably by the ISW and velocity contributions to the temperature
fluctuations in the CMB.  In Ref.~\cite{aurich}, it was found that the matching circles of a Poincar\'{e} 
dodecahedron universe cannot be found by current CMB searches because of this degradation.  In
Ref.~\cite{steiner}, the search included filtering and combines sets of expected circles, but disregarded
the $m$ weighting in the $S$ statistic, and provides no false positive thresholds.
As we have shown, the use of power spectrum filtering, the combination search, and the proper $S$ 
statistic that includes $m$ weighting can indeed probe beyond contaminating 
factors to determine if there are matching circles in the CMB. Applying these techniques to the
putative Poincar\'{e} Dodecahedral model produced no statistically significant sets of matching
circles. Since the smallest matching circles in our simulated sky have radii of $16.4^\circ$,
we have not been able to directly establish a lower limit on the angular radii that can be
probed by our combination search. However, we can extrapolate from the results shown in Figures
\ref{filter_fig} and \ref{combo_fig} to put the lower limit at $\alpha \simeq 5^\circ$.
This effectively rules out the Poincar\'{e} Dodecahedral model as an interesting shape for the Universe.

\section{Conclusion}\label{conclude}

	The claim that the topology of our Universe had been found to be that of the Poincar\'{e} 
dodecahedron does not stand up under scrutiny.  The signature found in Ref.~\cite{roukema} 
disappears when one uses the proper $S$ statistic and considers the false positive threshold.   
	While the shape of our Universe remains a mystery, the matching circles test can be used
to place a  lower bound on the size of the Universe.  The previous limit of 24 Gpc~\cite{cornish} can
be extended by about  10\% using filtering of the WMAP power spectrum. A full search with optimal filtering
is now underway.




\begin{thebibliography}{99}

\bibitem{stark} N.J. Cornish, D.N. Spergel \& G.D. Starkman, Class. Quantum Grav. {\bf 15}, 2657 (1998).

\bibitem{wmap} C.L. Bennett \emph {et al.}, Astrophys. J. Suppl. {\bf 148}, 1 (2003)

\bibitem{spergel} D.N. Spergel \emph {et al.}, Astrophys. J. Suppl. {\bf 148}, 175 (2003) 

\bibitem{cornish} N.J. Cornish, D.N. Spergel, G.D. Starkman \& E. Komatsu, 
Phys. Rev. Letters {\bf 92},  (2004).

\bibitem{jeff2} J. R. Weeks, Mod. Phys. Lett. A{\bf 18} 2099 (2003).

\bibitem{reza} B.~Mota, G.~I.~Gomero, M.~J.~Reboucas \& R.~Tavakol, Class. Quant. Grav. {\bf 21}, 3361 (2004).

\bibitem{gorski} J. Delabrouille, K.M. Gorski \& E. Hivon, preprint astro-ph/9710349 (1997).

\bibitem{roukema} B.F. Roukema, B. Lew, M. Cechowska, A. Marecki \& S. Bajtlik, 
Astro. \& Astrophy. {\bf }, (2004).

\bibitem{weeks} J-P. Luminet, J.R. Weeks, A. Riazuelo, R. Lehoucq, and J.-P. Uzan, Nature (London)
{\bf 425}, 593 (2003).

\bibitem{hu} W. Hu, N. Sugiyama \& J. Silk, Nature {\bf 386}, 37 (1997).

\bibitem{aurich} R. Aurich, S. Lustig \& F. Steiner, Class. Quantum Grav. {\bf 22}, 3443 (2005)

\bibitem{steiner} R. Aurich, S. Lustig \& F. Steiner, preprint astro-ph/0510847 (2005)

\bibitem{then} H. Then, preprint astro-ph/0511726 (2005).

\end{thebibliography}
\end{document}